\documentclass[12pt]{article}
\usepackage{amsfonts}
\newcommand{\stw}{{ }}
\newcommand{\text}{\rm}

\begin{document}

\title{\textbf{On the Renormalizability of Noncommutative U(1)
Gauge Theory - an Algebraic Approach}}
\author{L. C. Q. Vilar$^{c}$, O.S. Ventura$^{a,b}$, D.G.Tedesco$^{c}$, V. E. R. Lemes$^{c}$
\footnote{mbschlee@terra.com.br, ozemar@ifes.edu.br, vitor@dft.if.uerj.br}\\
{ \small \em $^a$Coordenadoria de F\'{\i}sica, Instituto Federal do Esp\'{\i}rito Santo},  \\
\small\em $ $Avenida Vit\'{o}ria 1729 - Jucutuquara, Vit\'{o}ria - ES, 29040 - 333, Brazil\\
\small\em $^b$Center for Theoretical Physics, Massachusetts Institute of Technology,\\
 \small\em77 Massachusetts Avenue, Cambridge, MA 02139, USA \\
\small\em $^c$Instituto de F\'\i sica, Universidade do Estado do Rio de Janeiro,\\
\small\em Rua S\~{a}o Francisco Xavier 524, Maracan\~{a}, Rio de Janeiro - RJ, 20550-013, Brazil \\}
\bigskip
\maketitle

\vspace{-1cm}
\begin{abstract}

We investigate the quantum effects of the nonlocal gauge invariant
operator $\frac{1}{{}{D}^{2}}{F}_{\mu \nu }\ast
\frac{1}{{}{D}^{2}}{F}^{\mu \nu }$ in the noncommutative $U(1)$
action and its consequences to the infrared sector of the theory.
Nonlocal operators of such kind were proposed to solve the
infrared problem of the noncommutative gauge theories evading the
questions on the explicit breaking of the Lorentz invariance. More
recently, a first step in the localization of this operator was
accomplished by means of the introduction of an extra tensorial
matter field, and the first loop analysis was carried out
$(Eur.Phys.J.\textbf{C62}:433-443,2009)$. We will complete this
localization avoiding the introduction of new degrees of freedom
beyond those of the original action by using only BRST doublets.
This will allow us to make a complete BRST algebraic study of the
renormalizability of the theory, following Zwanziger's method of
localization of nonlocal operators in QFT.

\end{abstract}
\setcounter{page}{0}\thispagestyle{empty}

\vfill\newpage\ \makeatother

\section{Introduction}

The year of 1999 witnessed two major developments in the
noncommutative quantum field theory program. In the first one,
Seiberg and Witten \cite{SW}, inspired by the previously known
result that the low energy limit of open strings could lead both
to a gauge theory defined on a noncommutative space as well as to
an usual commutative gauge theory, depending only on gauge choices,
announced the existence of what became called the Seiberg-Witten
map between noncommutative and commutative gauge theories.This
achievement was then fully tested and confirmed by several authors
both in the general structure of gauge transformations as in
specific examples of gauge theories (we make a short list of
references which is far from being complete
\cite{Asa,Grandi,Liu,pqs,0106188,Brandt,Yang,nosso,nosso2,Martin}).It also opened
a window to an alternative approach
to the quantum properties of the noncommutative theories.

The second development just revealed the kind of difficulties one
has to face when tackling the renormalization of  field theories
in the noncommutative space. An intrinsic mixing between high and
low energy scales was associated to the noncommutativity of
space-time, generating divergences which in the general case make
these theories not renormalizable as they stand \cite{Minw}, the
case of noncommutative gauge theories being no exception
\cite{Suss}. Recently, it was finally understood that this
infrared/ultraviolet (IR/UV) mix is still present even after a
Seiberg-Witten map \cite{You}, showing that the commutative
theories generated by their noncommutative counterparts suffer
from the same nonrenormalizability.

It took some time until the first proposal appeared in order to
cure a noncommutative scalar theory from this IR divergence
\cite{0401128}. The basic idea was to alter the free propagator of
the theory through the introduction of an harmonic potential, then
changing its low energy behavior.  This in fact made the theory
convergent in the infrared region, but at the cost of explicitly
breaking translation invariance. In \cite{0802.0791} this problem
was circumvented now by the introduction of a nonlocal term, again
assuring that the IR/UV mixing would be cured for a scalar theory.
Soon, this proposal was generalized to the case of a
noncommutative gauge theory \cite{0804.1914}. The main idea was
still the same, to change the low energy pattern of the theory,
and this was obtained through the introduction of a nonlocal term
one more time. The practical effect of this term is to modify the
free propagator of the gauge field, which acquires a $1/k^{4}$
pole, consistently defined in Euclidean space-time. This is how
the infrared regime of the theory gets modified. Again, we still
have a problem with this approach in the way that it was presented
up to this point, as the nonlocality is not adequate to match the
requisites of the Quantum Action Principle (QAP) \cite{livro} here
taken as valid in the noncommutative space (even though the
Quantun Action Principle has no proof of validity in the
noncommutative environment, its use became standard after the
results of \cite{martin99,maggiore,maggiore1,maggiore2}; we will
have more to say about this in the section IV.1). The way out
would be to find an equivalent local action meeting the same
properties of this previous one. So, the quantum study of such
theory had to wait until more recently, when a way to localize
this nonlocal action was found. Then a one-loop analysis was
finally carried out \cite{0901.1681}. This was an important
achievement, but once more there is an undesirable feature: the
introduction of an extra field in the theory, creating extra
degrees of freedom not present in the original noncommutative
gauge theory.  A natural question would be to ask if this is an
unavoidable price to be paid in order to have a possibly
renormalizable noncommutative theory with gauge interactions.

Our intention here will be to present an alternative scenario of
localization, pathing the way to a renormalizable noncommutative
gauge field theory, but avoiding to introduce any extra degree of
freedom.

In Section 2 we present the nonlocal action, its localization via
doublet fields and the resulting BRST symmetry. In Section 3 the
equations compatible with the Quantum Action Principle are
derived. Section 4 is dedicated to the analysis of the quantum
stability of the theory. In this section we pay special attention
to possible UV quantum corretions that can spoil the IR
renormalizability of the two poit function. The definitive form of
the propagator is finally obtained, showing a modification from
the classical starting one. In the final section we show our
conclusion.

\section{BRST in Euclidean space}

The nonlocal action that we will study is

\begin{equation}
S_{NL}=\int d^{4}x \lbrace \frac{1}{4}{ }{F}_{\mu \nu }\ast {F}^{\mu \nu }
+ \gamma^{4} \frac{1}{4}\frac{1}{{}{D}^{2}}{ }{F}_{\mu \nu }\ast \frac{1}{{}{D}^{2}}{F}^{\mu \nu } \rbrace . \label{snl}
\end{equation}
We are assuming an Euclidian signature for the space-time and an Abelian gauge group, with

 \begin{equation}
{ }{F}_{\mu \nu } = \partial_{\mu}{ }{A}_{\nu}- \partial_{\nu}{ }{A}_{\mu}
- ig [{ }{A}_{\mu} \stackrel{*}{,} { }{A}_{\nu}]  ,\;\;\;\;\;  {}{D}_{\mu} = \partial_{\mu} + ig [\;\;  \stackrel{*}{,} {}{A}_{\mu}] .\label{curv}
\end{equation}
The commutator of two coordinates is $[x^{\mu},x^{\nu}] = i\Theta^{\mu\nu} $, where
\begin{displaymath}
\label{theta}
\Theta= \left(
\begin{array}{llll}
   0 &\theta & 0 & 0\\
   -\theta & 0  & 0 & 0\\
   0&0 & 0 & \theta\\
   0& 0 & -\theta & 0
  \end{array} \right)
\end{displaymath}
and $\theta$ is the noncommutativity parameter \cite{0802.0791}.

This action gives to the gauge field propagator a more adequate behavior in the infrared for the noncommutative space

\begin{equation}
\left\langle
A(k)_\mu A_\nu(-k)\right\rangle = \left(\delta_{\mu\nu}
-\frac{k_\mu k_\nu}{k^2}\right)\frac{k^2}{k^{4} + \gamma^{4}}\;.
\label{ncmaxprop}
\end{equation}
As pointed out in \cite{0804.1914} and \cite{0901.1681}, the infrared behavior of this kind of propagator decouples the ultraviolet and infrared regimes, and, then, the action (\ref{snl}) is a good candidate to generate a coherent quantum gauge theory in noncommutative space, without the IR/UV mix.

The action $S_{NL}$ can be localized introducing a set of auxiliary tensorial fields. We use two pairs of
complex conjugated fields ${ }{\overline{B}}_{\mu \nu },{ }{B}_{\mu \nu };{ }{\overline{\chi}}_{\mu \nu },{ }{\chi}_{\mu \nu }$.
We will see that only with such structure one can hope to get rid of the unwanted extra degrees of freedom. Anyway, the action

\begin{eqnarray}
S_{LO} &=& S_{0} + S_{break}, \nonumber  \\
S_{0} &=& \int d^{4}x \lbrace
\frac{1}{4}{ }{F}_{\mu\nu}\ast{}{F}^{\mu \nu }
+ \stw{\overline{\chi}}_{\mu\nu}\ast D^{2}\stw{B}^{\mu\nu}+\stw{\overline{B}}_{\mu\nu}\ast D^{2}\stw{\chi}^{\mu\nu}
+\gamma^{2}\stw{\overline{\chi}}_{\mu\nu}\ast\stw{\chi}^{\mu\nu} \rbrace , \nonumber  \\
S_{break} &=& \int d^{4}x \lbrace - i\frac{\gamma}{2}\stw{B}_{\mu\nu}\ast \stw{F}^{\mu\nu}
+ i\frac{\gamma}{2}\stw{\overline{B}}_{\mu\nu}\ast \stw{F}^{\mu\nu} \rbrace , \label{inicio}
\end{eqnarray}
although representing the nonlocal operator of (\ref{snl}) in a localized form, still presents the problem that new degrees of freedom are being introduced by
the auxiliary fields.  This makes the physics content of the theory described by (\ref{inicio}) different from that of a noncommutative $U(1)$ theory.

This problem can be solved by associating a ghost for each tensorial field introduced, in
a way that a BRST structure of quartets will appear. This possibility of eliminating the extra degrees is the main reason for our choice of localization, as other attempts fail at this point. The action which attains this aim is

\begin{eqnarray}
S_{LO+G} &=& S_{0 +G} + S_{break} \nonumber  \\
S_{0 +G} &=& \int d^{4}x \lbrace \frac{1}{4}{ }{F}_{\mu\nu}\ast{}{F}^{\mu \nu }
+ \stw{\overline{\chi}}_{\mu\nu}\ast D^{2}\stw{B}^{\mu\nu}+\stw{\overline{B}}_{\mu\nu}\ast D^{2}\stw{\chi}^{\mu\nu} \nonumber  \\
&+&\gamma^{2}\stw{\overline{\chi}}_{\mu\nu}\ast\stw{\chi}^{\mu\nu}
- \stw{\overline{\psi}}_{\mu\nu}\ast \stw{D}^{2}\stw{\xi}^{\mu\nu}
- \stw{\overline{\xi}}_{\mu\nu}\ast \stw{D}^{2}\stw{\psi}^{\mu\nu}
-\gamma^{2}\stw{\overline{\psi}}_{\mu\nu}\ast\stw{\psi}^{\mu\nu} \rbrace .\label{localghost}
\end{eqnarray}

The action $S_{0 +G}$ is left invariant by the set of BRST transformations

\begin{eqnarray}
{}{s}{}{A}_{\mu}&=&-{}{D}_{\mu}{ }{c}
\mathrm{{\ },} \,\,\, { }{s}{ }{c} =-\frac{ig}{2}\lbrace{ }{c}\stackrel{*}{,}
{ }{c}\rbrace\mathrm{{\ },} \label{setbrs} \nonumber \\
{}{s}{ }{\overline{c}}&=&i{ }{b}\mathrm{{\ },}\,\,\,
{}{s}{ }{b}=0  \mathrm{{\ },}\nonumber \\
{}{s}{ }{F}_{\mu\nu}&=&-ig[ { }{c}%
\stackrel{*}{,}{ }{F}_{\mu \nu }] \mathrm{{\ },}  \nonumber \\
{}{s}{ }{\overline{\xi}}_{\mu \nu }&=&{ }{\overline{B}}_{\mu\nu}-ig\lbrace { }{c}%
\stackrel{*}{,}{ }{\overline{\xi}}_{\mu \nu }\rbrace \mathrm{{\ },}\,\,\,
{}{s}{ }{\overline{B}}_{\mu\nu}=-ig[ { }{c}%
\stackrel{*}{,}{ }{\overline{B}}_{\mu \nu }] \mathrm{{\ },}\nonumber \\
{}{s}{ }{\overline{\psi}}_{\mu \nu }&=&{ }{\overline{\chi}}_{\mu\nu}-ig\lbrace { }{c}%
\stackrel{*}{,}{ }{\overline{\psi}}_{\mu \nu }\rbrace \mathrm{{\ },}\,\,\,
{}{s}{ }{\overline{\chi}}_{\mu\nu}=-ig[ { }{c}%
\stackrel{*}{,}{ }{\overline{\chi}}_{\mu \nu }]\mathrm{{\ },} \nonumber \\
{ }{s}{ }{B}_{\mu \nu }&=&{ }{\xi}_{\mu\nu}-ig[ { }{c}%
\stackrel{*}{,}{ }{B}_{\mu \nu }] \mathrm{{\ },}\,\,\,
{}{s}{ }{\xi}_{\mu\nu}=-ig\lbrace { }{c}%
\stackrel{*}{,}{ }{\xi}_{\mu \nu }\rbrace \mathrm{{\ },}\nonumber \\
{ }{s}{ }{\chi}_{\mu \nu }&=&{ }{\psi}_{\mu\nu}-ig[ { }{c}%
\stackrel{*}{,}{ }{\chi}_{\mu \nu }] \mathrm{{\ },}\,\,\,
{}{s}{ }{\psi}_{\mu\nu}=-ig\lbrace { }{c}%
\stackrel{*}{,}{ }{\psi}_{\mu \nu }\rbrace \mathrm{{\ },}\label{BRST}\nonumber \\
\end{eqnarray}
where one can see the formation of a double quartet structure.
This is an important point to highlight here: the structure of 2
quartets is essential for the localization process. A possible
localization with only one quartet implies the use of the operator
$D^{2}{D}^{2}$ or other equivalent operator with $4$ derivatives,
and it is clear that this option leads to many nonrenormalizable
vertices for the localizing fields. These vertices carry large
momentum as expected by a theory that uses a field with canonical
dimension 1 and ultraviolet dimension 0 \cite{0903.4811}. In a commutative theory,
this fact certainly destroys the renormalizability, but a deeper
analysis is required in the case of noncommutative theories
due to the not very well known structure of the UV/IR mix and the
possibility of softening of divergences \cite{viena}. For such
reasons we decided to use $2$ quartets.

 The action $S_{0 +G}$ can then be written as
\begin{eqnarray}
S_{0 +G} &=& \int d^{4}x \lbrace \frac{1}{4}{ }{F}_{\mu\nu}\ast{}{F}^{\mu \nu } \rbrace + s\Delta^{-1}   ,  \nonumber \\
\Delta^{-1}&=& \int d^{4}x \lbrace  { }{\overline{\psi}}_{\mu \nu }\ast D^{2}\stw{B}^{\mu\nu}
+ \stw{\overline{\xi}}_{\mu\nu}\ast D^{2}\stw{\chi}^{\mu\nu} + \gamma^{2}{ }{\overline{\psi}}_{\mu \nu }\stw{\chi}^{\mu\nu} \rbrace . \label{delta}
\end{eqnarray}
Oncemore we notice that the physical degrees of freedom of the noncommutative $U(1)$ theory are being preserved.
In our localized action (\ref{localghost}) there is still a piece to be analyzed. The $S_{break}$ sector of the action
is not left invariant by the BRST transformations (\ref{BRST}). This is the element that will bring a new physics
to the pure U(1) case. It is BRST transformed into
\begin{eqnarray}
s S_{break} &=& \int d^{4}x \lbrace - i\frac{\gamma}{2}{\xi}_{\mu\nu}\ast \stw{F}^{\mu\nu} \rbrace  .  \label{quebra}
\end{eqnarray}
From this point on we will assume that the Moyal product is rigid under quantum corrections. In the noncommutative space, the Moyal structure is intimately related to the gauge symmetry, and one cannot modify the first without damaging the latter.
This can also be infered from the fact that the only nontrivial cocycles of the BRST cohomology of gauge theories involve exclusively
terms constructed with the field strength and covariant derivatives at the level of the counterterms in the study of the quantum stability of
the gauge action \cite{livro}. Naturally, in the noncommutative space, there is room for higher dimensional terms built explicitly with $\theta$, field strengths and covariant derivatives, invariant and nontrivial in the BRST sense, which are not present at the original action. This is also
seen by the method of consistent deformations of \cite{Henneaux} applied to the present case of noncommutative deformations of
Maxwell theory \cite{0106188}. It is the Lorentz structure of the vertices of the theory together with gauge invariance which prohibits such counterterms. In \cite{0707.3681}, explicit calculations
 in noncommutative Chern-Simons theory showed these properties. Then, although the presence of $F_{\mu\nu}$ in (\ref{quebra})
 implies an infinite series of terms, the rigidity of the Moyal product determines that $F_{\mu\nu}$ is renormalized as a whole.
 This allows us to understand the breaking in (\ref{quebra}) in a way analogous to that of a soft breaking in a commutative theory
 (one can see that in zero $\theta$ order, this breaking is undoubtedly a soft breaking).  The treatment of softly broken
 theories was recently formalized in \cite{baulieu}. We will need to study the renormalization of the theory together with the renormalization of the breaking itself. This
is done by introducing a set of sources in a BRST doublet in such a way that the physical action is obtained when we set the
sources to their physical values:

\begin{eqnarray}
S_{break} &=& S_{source}\Big|_{\mathrm{phys}}\nonumber \\
S_{source} &=& \int d^{4}x (
\stw{\overline{J}}_{\mu\nu\alpha\beta} \ast \lbrace {B}^{\mu\nu}\stackrel{*}{,}
\stw{F}^{\alpha\beta}\rbrace +
\stw{J}_{\mu\nu\alpha\beta}\ast \lbrace\overline{B}^{\mu\nu}\stackrel{*}{,}
\stw{F}^{\alpha\beta}\rbrace -
\stw{\overline{Q}}_{\mu\nu\alpha\beta} \ast \lbrace\xi^{\mu\nu}\stackrel{*}{,}
\stw{F}^{\alpha\beta} \rbrace ) ,\label{source}
\end{eqnarray}
where by $\Big|_{\mathrm{phys}}$ we mean that in this limit the sources attain their physical values,
\begin{eqnarray}
\stw{J}_{\mu\nu\alpha\beta}|&=&\frac{i}{8}\gamma(\delta_{\mu\alpha}\delta_{\beta\nu}-
\delta_{\mu\beta}\delta_{\alpha\nu})\mathrm{{\ },}\,\,\,
\stw{\overline{J}}_{\mu\nu\alpha\beta}|=
-\frac{i}{8}\gamma(\delta_{\mu\alpha}\delta_{\beta\nu}- \delta_{\mu\beta}\delta_{\alpha\nu}), \nonumber \\
\stw{Q}_{\mu\nu\alpha\beta}|&=& 0\mathrm{{\ },}\,\,\,
\stw{\overline{Q}}_{\mu\nu\alpha\beta}|=0. \label{pval}
\end{eqnarray}
The BRST transformation of the sources,
\begin{eqnarray}
{}{s} \stw{Q}_{\mu\nu\alpha\beta} &=& \stw{J}_{\mu\nu\alpha\beta}-ig\lbrace { }{c}%
\stackrel{*}{,}{ }{Q}_{\mu\nu\alpha\beta }\rbrace
\mathrm{{\ },}{}\,\,\, {s} \stw{J}_{\mu\nu\alpha\beta} = -ig[ { }{c}%
\stackrel{*}{,}{ }{J}_{\mu\nu\alpha\beta }], \label{fontes} \\
\stw{s}\stw{\overline{Q}}_{\mu\nu\alpha\beta} &=&\stw{\overline{J}}_{\mu\nu\alpha\beta}
-ig\lbrace { }{c}%
\stackrel{*}{,}{ }{\overline{Q}}_{\mu\nu\alpha\beta }\rbrace
\mathrm{{\ },}{}\,\,\, {s} \stw{\overline{J}}_{\mu\nu\alpha\beta}=-ig[ { }{c}%
\stackrel{*}{,}{ }{\overline{J}}_{\mu\nu\alpha\beta }],
\end{eqnarray}
shows the doublet structure that we have already mentioned. The
action (\ref{source}) is now easily seen as an exact BRST
variation, and the process altogether is a kind of an immersion of
the original theory inside this more general one. Following this
reasoning, we can now also rewrite the mass term
$\gamma^{2}\overline{\chi}_{\mu\nu}\chi_{\mu\nu}-\gamma^{2}\overline{\psi}_{\mu\nu}\psi_{\mu\nu}$
in such a way that the mass parameter $\gamma^{2}$ only appears in
the theory after taking the physical values for the sources $J,
\overline{J}, Q, \overline{Q}$. This approach turns it easier to
note that before this process only the original degrees of freedom
coming from the gauge field $A_{\mu}$ are present in the action.

The last steps needed for the BRST quantization are the definition of a gauge fixing, which we take as the noncommutative Landau gauge fixing,
\begin{equation}
 S_{gf} =\int d^{4}x \lbrace i{ }{b}\ast\partial_{\mu}{ }{A}^{\mu}
+{ }{\overline{c}}\ast\partial^{\mu}{ }{D}_{\mu}{ }{c} \rbrace .
\end{equation}
And finally, a set of Slavnov sources ${ \Omega , L, \overline{u}, u, \overline{v}, v, \overline{P}, P, \overline{R}, R, \overline{M}, M, \overline{N}, N }$  are introduced in the action coupled to the nonlinear BRST transformations of the fields $A,c,\xi , \overline{\xi},B , \overline{B}, \psi , \overline{\psi}, \chi , \overline{\chi} $ and sources $Q , \overline{Q},J , \overline{J}$ respectively.

The complete invariant action can then be written as

\begin{eqnarray}
\Sigma &=&\int d^{4}x \lbrace  \frac{1}{4}{ }{F}_{\mu \nu }\ast{ }{F}^{\mu \nu } + i{ }{b}\ast\partial_{\mu}{ }{A}^{\mu}
+{ }{\overline{c}}\ast\partial^{\mu}{ }{D}_{\mu}{ }{c}
+ \stw{\overline{\chi}}_{\mu\nu}\ast D^{2}\stw{B}^{\mu\nu}+\stw{\overline{B}}_{\mu\nu}\ast D^{2}\stw{\chi}^{\mu\nu} \nonumber \\
&+& \stw{\overline{J}}_{\mu\nu\alpha\beta} \ast \lbrace
{B}^{\mu\nu}\stackrel{*}{,} \stw{F}^{\alpha\beta}\rbrace +
\stw{J}_{\mu\nu\alpha\beta}\ast
\lbrace\overline{B}^{\mu\nu}\stackrel{*}{,}
\stw{F}^{\alpha\beta}\rbrace - \stw{\overline{\psi}}_{\mu\nu}\ast
\stw{D}^{2}\stw{\xi}^{\mu\nu}
- \stw{\overline{\xi}}_{\mu\nu}\ast \stw{D}^{2}\stw{\psi}^{\mu\nu}\nonumber \\
&+& \frac{2}{3} \lbrace
\overline{J}_{\alpha\beta\sigma\lambda}\stackrel{*}{,}J^{\alpha\beta\sigma\lambda}\rbrace
\lbrace
\overline{\chi}_{\mu\nu}\stackrel{*}{,}\chi^{\mu\nu}\rbrace -
\frac{2}{3}\lbrace
\overline{J}_{\alpha\beta\sigma\lambda}\stackrel{*}{,}J^{\alpha\beta\sigma\lambda}\rbrace
[ \overline{\psi}_{\mu\nu}\stackrel{*}{,}\psi^{\mu\nu} ] \nonumber \\
&-&\stw{\overline{Q}}_{\mu\nu\alpha\beta} \ast
\lbrace\xi^{\mu\nu}\stackrel{*}{,} \stw{F}^{\alpha\beta} \rbrace -
\stw{\Omega}_{\mu}\ast \stw{D}^{\mu}\stw{c} - \frac{i}{2}L\ast
g\lbrace{ }{c}\stackrel{*}{,}{ }{c}\rbrace \nonumber \\
&-& i\stw{\overline{u}}^{\mu\nu}\ast g\lbrace \stw{c}\stackrel{*}{,}
\stw{\xi}_{\mu\nu}\rbrace
+ \stw{u}^{\mu\nu}\ast({ }{\overline{B}}_{\mu\nu}-ig\lbrace { }{c}%
\stackrel{*}{,}{ }{\overline{\xi}}_{\mu \nu }\rbrace)\nonumber \\
&+& \stw{\overline{v}}^{\mu\nu}\ast({ }{\xi}_{\mu\nu}-ig[ { }{c}%
\stackrel{*}{,}{ }{B}_{\mu \nu }]) - i \stw{v}^{\mu\nu}\ast g [ { }{c}%
\stackrel{*}{,}{ }{\overline{B}}_{\mu \nu }]\nonumber \\
&-&i\stw{\overline{P}}^{\mu\nu}\ast g\lbrace { }{c}%
\stackrel{*}{,}{ }{\psi}_{\mu \nu }\rbrace
+\stw{P}^{\mu\nu}\ast(\stw{\overline{\chi}}_{\mu \nu }-ig\lbrace { }{c}%
\stackrel{*}{,}{ }{\overline{\psi}}_{\mu \nu }\rbrace )\nonumber \\
&+& \stw{\overline{R}}^{\mu\nu}\ast(\stw{{\psi}}_{\mu \nu }-ig[ { }{c}%
\stackrel{*}{,}{ }{{\chi}}_{\mu \nu }] ) -iR^{\mu\nu}\ast g[ { }{c}%
\stackrel{*}{,}{ }{{\overline{\chi}}}_{\mu \nu }] \nonumber \\
&+& \overline{M}^{\mu\nu\alpha\beta}\ast (\stw{J}_{\mu\nu\alpha\beta}-ig\lbrace { }{c}%
\stackrel{*}{,}{ }{Q}_{\mu\nu\alpha\beta }\rbrace )
+ M^{\mu\nu\alpha\beta}\ast(\overline{J}_{\mu\nu\alpha\beta}-ig\lbrace { }{c}%
\stackrel{*}{,}{ }\overline{Q}_{\mu\nu\alpha\beta }\rbrace ) \nonumber \\
&-& i\overline{N}^{\mu\nu\alpha\beta}\ast g[ { }{c}%
\stackrel{*}{,}{ }{J}_{\mu\nu\alpha\beta }]
-i N^{\mu\nu\alpha\beta}\ast g[ { }{c}%
\stackrel{*}{,}{ }\overline{J}_{\mu\nu\alpha\beta }]
\,\,\,\,\, \rbrace ,
\label{acao}
\end{eqnarray}
and it is ready for the BRST analysis.

\section{Equations Compatible with the Quantum Action Principle}

In this section we will present several symmetries compatible with the QAP, which will be useful in the BRST renormalization procedure. First we have the traditional Ward identities present in usual gauge theories:

\begin{itemize}

\item Slavnov Taylor
\begin{eqnarray}
S(\Sigma ) &=& \int d^{4}x \lbrace \frac{\delta\Sigma}{\delta\stw{A}_{\mu}}\frac{\delta\Sigma}{\delta\stw{\Omega}^{\mu}}
+ \frac{\delta\Sigma}{\delta\stw{c}}\frac{\delta\Sigma}{\delta\stw{L}}
+ i\stw{b}\frac{\delta\Sigma}{\delta\stw{\overline{c}}}
+\frac{\delta\Sigma}{\delta\stw{\overline{u}}^{\mu\nu}}
\frac{\delta\Sigma}{\delta\stw{\xi}_{\mu\nu}}
+\frac{\delta\Sigma}{\delta\stw{u}^{\mu\nu}}
\frac{\delta\Sigma}{\delta\stw{\overline{\xi}}_{\mu\nu}}\nonumber \\
&+& \frac{\delta\Sigma}{\delta\stw{\overline{v}}^{\mu\nu}}
\frac{\delta\Sigma}{\delta\stw{B}_{\mu\nu}}
+ \frac{\delta\Sigma}{\delta\stw{v}^{\mu\nu}}
\frac{\delta\Sigma}{\delta\stw{\overline{B}}_{\mu\nu}}
+ \frac{\delta\Sigma}{\delta\stw{\overline{P}}^{\mu\nu}}
\frac{\delta\Sigma}{\delta\stw{\psi}_{\mu\nu}}
+ \frac{\delta\Sigma}{\delta\stw{P}^{\mu\nu}}
\frac{\delta\Sigma}{\delta\stw{\overline{\psi}}_{\mu\nu}} \nonumber \\
&+& \frac{\delta\Sigma}{\delta\stw{\overline{R}}^{\mu\nu}}
\frac{\delta\Sigma}{\delta\stw{\chi}_{\mu\nu}}
+ \frac{\delta\Sigma}{\delta\stw{R}^{\mu\nu}}
\frac{\delta\Sigma}{\delta\stw{\overline{\chi}}_{\mu\nu}}
+ \frac{\delta\Sigma}{\delta J^{\sigma\lambda\alpha\beta}}\frac{\delta\Sigma}{\delta \overline{N}_{\sigma\lambda\alpha\beta}} + \frac{\delta\Sigma}{\delta \overline{J}^{\sigma\lambda\alpha\beta}}\frac{\delta\Sigma}{\delta N_{\sigma\lambda\alpha\beta}} \nonumber \\
&+& \frac{\delta\Sigma}{\delta Q^{\sigma\lambda\alpha\beta}}\frac{\delta\Sigma}{\delta \overline{M}_{\sigma\lambda\alpha\beta}} + \frac{\delta\Sigma}{\delta \overline{Q}^{\sigma\lambda\alpha\beta}}\frac{\delta\Sigma}{\delta M_{\sigma\lambda\alpha\beta}}
\rbrace ,\label{slavnov}
\end{eqnarray}

\item Lagrange multiplier and antighost equation
\begin{equation}
\frac{\delta\Sigma}{\delta \stw{b}}= i\partial^{\mu}\stw{A}_{\mu\nu}\mathrm{{\ },} \,\,\,\, \partial_{\mu}\frac{\delta\Sigma}{\delta\stw{\Omega}_{\mu}}
+\frac{\delta\Sigma}{\delta\stw{\overline{c}}}=0 , \label{equatcb} \\
\end{equation}

\item Ghost equation
\begin{equation}
\mathcal{G}\Sigma = \int d^{4}x
\frac{\delta\Sigma}{\delta\stw{c}} = 0,
\label{ghosteq}
\end{equation}

\item $SL(2,R)$ equation
\begin{equation}
D(\Sigma) = \int d^{4}x \lbrace c\frac{\delta\Sigma}{\delta\stw{\overline{c}}} - i \frac{\delta\Sigma}{\delta b}\frac{\delta\Sigma}{\delta L} \rbrace = 0.
\label{sl2r}
\end{equation}

\end{itemize}

It is important to emphasize here that, due to the Moyal structure,
the possible breaking terms are vanishing when integrated.

Now, due to the fact that all couplings are derivative in the noncommutative $U(1)$ theory, we also have  integrated equations of motion,

\begin{eqnarray}
\int d^{4}x
\frac{\delta\Sigma}{\delta\stw{\overline{\chi}}^{\mu\nu}} &=& \int
d^{4}x (\frac{4}{3}\lbrace
\overline{J}_{\alpha\beta\sigma\lambda}\stackrel{*}{,}J^{\alpha\beta\sigma\lambda}\rbrace
\stw{\chi}_{\mu\nu} + \stw{P}_{\mu\nu}),\nonumber \\
\int d^{4}x \frac{\delta\Sigma}{\delta\stw{\chi}^{\mu\nu}} &=&
\frac{4}{3}\int d^{4}x (\lbrace
\overline{J}_{\alpha\beta\sigma\lambda}\stackrel{*}{,}J^{\alpha\beta\sigma\lambda}\rbrace\stw{\overline{\chi}}_{\mu\nu} ),\nonumber \\
\int d^{4}x
\frac{\delta\Sigma}{\delta\stw{\overline{\psi}}^{\mu\nu}} &=&
-\frac{4}{3}\int d^{4}x (\lbrace
\overline{J}_{\alpha\beta\sigma\lambda}\stackrel{*}{,}J^{\alpha\beta\sigma\lambda}\rbrace\stw{\psi}_{\mu\nu} ),\nonumber \\
\int d^{4}x \frac{\delta\Sigma}{\delta\stw{\psi}^{\mu\nu}} &=&
\int d^{4}x
(\frac{4}{3}\lbrace\overline{J}_{\alpha\beta\sigma\lambda}\stackrel{*}{,}J^{\alpha\beta\sigma\lambda}\rbrace\stw{\overline{\psi}}_{\mu\nu}
+ \stw{\overline{R}}_{\mu\nu}), \nonumber \\
\int d^{4}x \frac{\delta\Sigma}{\delta\overline{\xi}^{\mu\nu}} &=& 0 .
\label{eqmovint}
\end{eqnarray}
These Ward identities will play a major role in the
renormalizability study that we will derive. Let us observe here
that these symmetries are only present in the $U(1)$ case, for the
general $U(N)$ theory has nonderivative interactions. The absence
of the Ward identities (\ref{eqmovint}) is the main reason why we
believe that in the nonabelian noncommutative case we need an
alternative way of approaching the IR/UV problem. Now, let us  go
back to the $U(1)$ case.

The final symmetries that we will list are the identity associated to the BRST doublet structure $U^{(1)}$,
\begin{eqnarray}
{U^{(1)}}_{\sigma\lambda\mu\nu}(\Sigma ) &=& \int d^{4}x
(\stw{\xi}_{\sigma\lambda}\frac{\delta\Sigma}{\delta\stw{B}_{\mu\nu}}
+
\stw{\overline{B}}_{\mu\nu}\frac{\delta\Sigma}{\delta\stw{\overline{\xi}}_{\sigma\lambda}}
+
\stw{\overline{\chi}}_{\mu\nu}\frac{\delta\Sigma}{\delta\stw{\overline{\psi}}_{\sigma\lambda}}+
\stw{\psi}_{\sigma\lambda}\frac{\delta\Sigma}{\delta\stw{\chi}_{\mu\nu}} \nonumber \\
&+&
\stw{\overline{J}}_{\mu\nu\alpha\beta}\frac{\delta\Sigma}{\delta\overline{Q}_{\sigma\lambda\alpha\beta}}
+ \stw{M}_{\sigma\lambda\alpha\beta}\frac{\delta\Sigma}{\delta N_{\mu\nu\alpha\beta}}+
\stw{J}_{\sigma\lambda\alpha\beta}\frac{\delta\Sigma}{\delta\stw{Q}_{\mu\nu\alpha\beta}}
+
\stw{\overline{M}}_{\mu\nu\alpha\beta}\frac{\delta\Sigma}{\delta\stw{\overline{N}_{\sigma\lambda\alpha\beta}}}
\nonumber \\
&+&
\stw{u}_{\sigma\lambda}\frac{\delta\Sigma}{\delta\stw{v}_{\mu\nu}}
+
 \stw{\overline{v}}_{\mu\nu}\frac{\delta\Sigma}{\delta\stw{\overline{u}}_{\sigma\lambda}}+
\stw{P}_{\sigma\lambda}\frac{\delta\Sigma}{\delta\stw{R}_{\mu\nu}}
+
\stw{\overline{R}}_{\mu\nu}\frac{\delta\Sigma}{\delta\stw{\overline{P}}_{\sigma\lambda}})
= 0 , \label{u1}
\end{eqnarray}
the linearly broken symmetries $U^{(0)}$ and $\widetilde{U}^{(0)}$,
\begin{eqnarray}
{U^{(0)}}_{\sigma\lambda\mu\nu}(\Sigma ) &=& - {\Theta^{(0)}}_{\sigma\lambda\mu\nu} \nonumber \\
{U^{(0)}}_{\sigma\lambda\mu\nu}(\Sigma ) &=& \int d^{4}x
(\stw{B}_{\sigma\lambda}\frac{\delta\Sigma}{\delta\stw{B}_{\mu\nu}}
-
\stw{\overline{B}}_{\mu\nu}\frac{\delta\Sigma}{\delta\stw{\overline{B}}_{\sigma\lambda}}
+
\stw{\chi}_{\sigma\lambda}\frac{\delta\Sigma}{\delta\stw{\chi}_{\mu\nu}}
-
\stw{\overline{\chi}}_{\mu\nu}\frac{\delta\Sigma}{\delta\stw{\overline{\chi}}_{\sigma\lambda}} \nonumber \\
&+&
\stw{J}_{\sigma\lambda\alpha\beta}\frac{\delta\Sigma}{\delta\stw{J}_{\mu\nu\alpha\beta}}
-
\stw{\overline{N}}_{\mu\nu\alpha\beta}\frac{\delta\Sigma}{\delta\stw{\overline{N}_{\sigma\lambda\alpha\beta}}}
- \overline{J}_{\mu\nu\alpha\beta}\frac{\delta\Sigma}{\overline{J}_{\sigma\lambda\alpha\beta}}
+ N_{\sigma\lambda\alpha\beta}\frac{\delta\Sigma}{N}_{\mu\nu\alpha\beta}
\nonumber \\
&+&
\stw{R}_{\sigma\lambda}\frac{\delta\Sigma}{\delta\stw{R}_{\mu\nu}}
-
\stw{\overline{R}}_{\mu\nu}\frac{\delta\Sigma}{\delta\stw{\overline{R}}_{\sigma\lambda}}
+
\stw{v}_{\sigma\lambda}\frac{\delta\Sigma}{\delta\stw{v}_{\mu\nu}}
-
\stw{\overline{v}}_{\mu\nu}\frac{\delta\Sigma}{\delta\stw{\overline{v}}_{\sigma\lambda}})
\nonumber \\
{\Theta^{(0)}}_{\sigma\lambda\mu\nu} &=& \int d^{4}x (
\stw{u}_{\sigma\lambda}\ast\stw{\overline{B}}_{\mu\nu} +
\stw{P}_{\sigma\lambda}\ast\stw{\overline{\chi}}_{\mu\nu} +
\stw{\overline{R}}_{\mu\nu}\ast\stw{\psi}_{\sigma\lambda} +
\stw{\overline{v}}_{\mu\nu}\ast\stw{\xi}_{\sigma\lambda}\nonumber \\ &+&
\overline{J}_{\mu\nu}{}^{\alpha\beta}\ast M_{\sigma\lambda\alpha\beta}-
\overline{M}_{\mu\nu}{}^{\alpha\beta}\ast J_{\sigma\lambda\alpha\beta}),
 \label{u0}
\end{eqnarray}

\begin{eqnarray}
{\widetilde{U}^{(0)}}_{\sigma\lambda\mu\nu}(\Sigma ) &=& {\Theta^{(0)}}_{\mu\nu\sigma\lambda} \nonumber \\
{\widetilde{U}^{(0)}}_{\sigma\lambda\mu\nu}(\Sigma ) &=& \int
d^{4}x
(\stw{\psi}_{\mu\nu}\frac{\delta\Sigma}{\delta\stw{\psi}_{\sigma\lambda}}
-
\stw{\overline{\psi}}_{\sigma\lambda}\frac{\delta\Sigma}{\delta\stw{\overline{\psi}}_{\mu\nu}}
+
\stw{\xi}_{\mu\nu}\frac{\delta\Sigma}{\delta\stw{\xi}_{\sigma\lambda}}
-
\stw{\overline{\xi}}_{\sigma\lambda}\frac{\delta\Sigma}{\delta\stw{\overline{\xi}}_{\mu\nu}} \nonumber \\
&-& \stw{\overline{Q}}_{\sigma\lambda\alpha\beta}\frac{\delta\Sigma}{\delta\stw{\overline{Q}}_{\mu\nu\alpha\beta}}
+ M_{\mu\nu\alpha\beta}\frac{\delta\Sigma}{\delta M_{\sigma\lambda\alpha\beta}}
+ Q_{\mu\nu\alpha\beta}\frac{\delta\Sigma}{\delta Q_{\sigma\lambda\alpha\beta}}
-\overline{M}_{\sigma\lambda\alpha\beta}\frac{\delta\Sigma}{\delta \overline{M}_{\mu\nu\alpha\beta}}
\nonumber \\
&+&
\stw{P}_{\mu\nu}\frac{\delta\Sigma}{\delta\stw{P}_{\sigma\lambda}}
-
\stw{\overline{P}}_{\sigma\lambda}\frac{\delta\Sigma}{\delta\stw{\overline{P}}_{\mu\nu}} +
\stw{u}_{\mu\nu}\frac{\delta\Sigma}{\delta\stw{u}_{\sigma\lambda}}
-
\stw{\overline{u}}_{\sigma\lambda}\frac{\delta\Sigma}{\delta\stw{\overline{u}}_{\mu\nu}}),
\label{tildeu0}
\end{eqnarray}
which together define the reality constraint on our action $\Sigma$ and an associated reality charge $Q$ for all the fields and sources of the theory,

\begin{equation}
 Q =   Tr U^{(0)} + Tr \widetilde{U}^{(0)},
\end{equation}
and finally the last two symmetries

\begin{equation}
{U^{(2)}}_{\sigma\lambda\mu\nu}(\Sigma ) =  \int d^{4}x \lbrace \psi_{\sigma\lambda}\frac{\delta\Sigma}{\delta \overline{\xi}^{\mu\nu}} + \psi_{\mu\nu}\frac{\delta\Sigma}{\delta \overline{\xi}^{\sigma\lambda}} - u_{\sigma\lambda}\frac{\delta\Sigma}{\delta \overline{P}^{\mu\nu}} - u_{\mu\nu}\frac{\delta\Sigma}{\delta \overline{P}^{\sigma\lambda}} \rbrace =0,\label{u2}
\end{equation}
and
\begin{eqnarray}
{\widetilde{U}^{(2)}}_{\sigma\lambda\mu\nu}(\Sigma ) &=&  \int d^{4}x \lbrace \psi_{\sigma\lambda}\frac{\delta\Sigma}{\delta \overline{\psi}^{\mu\nu}} + \psi_{\mu\nu}\frac{\delta\Sigma}{\delta \overline{\psi}^{\sigma\lambda}} +
\xi_{\sigma\lambda}\frac{\delta\Sigma}{\delta \overline{\xi}^{\mu\nu}} + \xi_{\mu\nu}\frac{\delta\Sigma}{\delta \overline{\xi}^{\sigma\lambda}} \nonumber \\
&-& u_{\sigma\lambda}\frac{\delta\Sigma}{\delta \overline{u}^{\mu\nu}} - u_{\mu\nu}\frac{\delta\Sigma}{\delta \overline{u}^{\sigma\lambda}}
- P_{\sigma\lambda}\frac{\delta\Sigma}{\delta \overline{P}^{\mu\nu}} - P_{\mu\nu}\frac{\delta\Sigma}{\delta \overline{P}^{\sigma\lambda}}
\rbrace =0.\label{tildeu2}
\end{eqnarray}

Let us already explain here that the tensorial nature of these symmetries
will be responsible for the fact that,in the cohomological
analysis that we will undertake, the only possible Lorentz indices contractions of the fields
$\chi
,\overline{\chi},B,\overline{B},\psi ,\overline{\psi},\xi
,\overline{\xi} $ and their sources obey the same structure present
in the action (\ref {acao}).

\section{Stability of the quantum action}
In order to study the stability of the quantum action let us start by presenting the quantum numbers of all fields and sources:

\begin{table}[h]
\centering
\begin{tabular}{|c|c|c|c|c|c|c|c|c|c|c|c|c|}
\hline
fields & $\stw{A}$ & $\stw{b}$ & $\stw{c}$ & $\stw{\overline{c}}$ & $\stw{\psi}$ & $\stw{\overline{\psi}}$ & $\stw{\xi}$ & $\stw{\overline{\xi}}$ & $\stw{\chi}$ & $\stw{\overline{\chi}}$ & $\stw{B}$ & $\stw{\overline{B}}$  \\
\hline
UV dimension & 1 & 2 & 0 & 2 & 1 & 1 & 1 & 1 & 1 & 1 & 1 & 1\\
Ghost number & 0 & 0 & 1 & $-1$ & 1 & $-1$ & 1 & $-1$ & 0 & 0 & 0 & 0 \\
Q charge & 0 & 0 & 0 & 0 & 1 & $-1$ & 1 & $-1$ & 1 & $-1$ & 1 & $-1$ \\
Statistics & co & co & an & an & an & an & an & an & co & co & co & co \\
\hline
\end{tabular}
\caption{Quantum numbers of the fields.}
\label{table1}
\end{table}

\begin{table}[h]
\centering
\begin{tabular}{|c|c|c|c|c|c|c|c|c|c|c|c|c|c|c|}
\hline
sources & $\stw{\Omega}$ & $\stw{L}$ & $\stw{J}$ & $\stw{\overline{J}}$ & $\stw{Q}$ & $\stw{\overline{Q}}$ & $\stw{u}$ & $\stw{\overline{u}}$ & $\stw{v}$ & $\stw{\overline{v}}$ & $\stw{P}$ & $\stw{\overline{P}}$  & $\stw{R}$ & $\stw{\overline{R}}$\\
\hline
UV dimension & 3 & 4 & 1 & 1 & 1 & 1 & 3 & 3 & 3 & 3 & 3 & 3 & 3 & 3\\
Ghost number & $-1$ & $-2$ & 0 & 0 & $-1$ & $-1$ & 0 & $-2$ & $-1$ & $-1$ & 0 & $-2$ & $-1$ & $-1$\\
Q charge & 0 & 0 & 1 & $-1$ & 1 & $-1$ & 1& $-1$ & 1 & $-1$ & 1 & $-1$ & 1 & $-1$ \\
Statistics & an & co & co & co & an & an & co & co & an & an & co & co & an & an \\
\hline
\end{tabular}
\caption{Quantum numbers of the sources.}
\label{table2}
\end{table}

\begin{table}[h]
\centering
\begin{tabular}{|c|c|c|c|c|}
\hline
sources & $M$ & $N$ & $\overline{M}$ & $\stw{\overline{N}}$  \\
\hline
UV dimension & 3 & 3 & 3 & 3 \\
Ghost number & 0 & $-1$ & 0 &  $-1$ \\
Q charge & 1 & 1 & $-1$ & $-1$ \\
Statistics & co & an & co & an  \\
\hline
\end{tabular}
\caption{Quantum numbers of the auxiliary sources}
\label{table3}
\end{table}
Once more, we call attention to the fact that in the stability
analysis of the quantum action it is necessary to take into
account not the canonical dimension but the ultraviolet dimension
of all fields. The use of canonical dimensions generally leads to an
incorrect cohomological analysis.

\subsection{The invariant counterterm}

In this section we will focus our attention on the posible UV
counterterms that can change the propagation behavior of the
classical theory. The original structure that is obtained from
(\ref{acao}) when the sources attain their physical values is
specially designed in order to incorporate the coefficient of
the IR singularity
appearing in the two point function of the noncommutative $U(1)$
theory \cite{0804.1914} (other singular IR contributions are not
addressed in this analysis \cite{martin2002}). Then, new UV
counterterms different from those already present in the starting
action (\ref{acao}) can be rather harmful to the delicate match at
the IR level. The search for such contributions is our main
interest here.

Before proceeding, we would like to spend a few words on the use
of the QAP in this noncommutative context. Let us recapitulate the
origin of the IR/UV mix. In general, Feynman graphs calculations
in noncommutative theories can be divided in planar and non-planar
contributions \cite{Minw}. The latter are those characterized by
the presence of a remaining phase inside the Feynman integrals.
This phase is responsible for the damping of the UV divergences,
which become naturally regularized. As this phase depends on the
external momenta (the phase disappears for vanishing external
momenta), the would be UV divergence is turned into an UV finite
but IR singular contribution. The introduction of these non-local
objects in the starting action is a possible mechanism that is
actually behind the reasoning leading to the proposal of the
action (\ref{snl}) to account for the two point function IR
singularity of the pure noncommutative $U(1)$ theory. On the other
side, the non-planar graph is accompanied by its planar
counterpart, when the phase becomes dependent only on the external
momenta, and, in this way, factorizes off the integral. In
general, the UV divergence of a planar graph is accompanied by the
non-planar singularity, generating the IR/UV mix. But the point is
that the $U(1)$ planar graphs, where the UV divergences are
generated, mimic the structure of a commutative theory inside the
integrals, with phase dependent coefficients restoring the
noncommutative vertices \cite{martin99}. This is what we meant by
the Moyal rigidity hipothesis in the introduction. Then, it is in
this sector of the noncommutative theory that the QAP seems to be
valid, with a power-counting bounding the possible UV
counterterms. The use of the QAP can then be a guidance to search
for possible IR singularities in the non-planar counterparts
associated to the planar UV divergent contributions. Finally, once
all IR singularities of a previous theory are stabilized, using
mechanisms as that in (\ref{snl}), the question if the new
non-local action developes new IR singularities can again be
answered using the QAP in the planar sector. If the QAP indicates
that this IR stabilized theory has no new UV divergent
contributions, we would be meeting a renormalization condition for
this final theory.
 In order to characterize any
invariant counterterm which can be added freely to all orders in
perturbation theory \cite{livro}, we perturb the classical action
$\Sigma$ by adding an arbitrary integrated local polynomial
$\Sigma^{count}$ of dimension up-bounded by four, vanishing ghost
number and Q charge. We demand that
$\Gamma=\Sigma+\epsilon\Sigma^{count}+O(\epsilon^2)$, where
$\epsilon$ is a small expansion parameter, satisfies the same Ward
identities as $\Sigma$. This requirement provides the following
constraints on the counterterm (for convenience of the reader, we
follow the same sequence of Ward identities of section(3)):

\begin{eqnarray}
{\cal B}_{\Sigma}\Sigma^{count}&=&0\;,\label{1}\\
\frac{\delta}{\delta \stw{b}}\Sigma^{count}&=&0\;,\label{2}\\
\partial_{\mu}\frac{\delta\Sigma^{count}}{\delta\stw{\Omega}_{\mu}}
+\frac{\delta\Sigma^{count}}{\delta\stw{\overline{c}}}&=& 0\;, \label{3}\\
D_{\Sigma}\Sigma^{count}&=&0\;,\label{4}\\
\mathcal{G}\Sigma^{count} &=& 0\;,      \label{5}\\
\int d^{4}x \frac{\delta\Sigma^{count}}{\delta\stw{\overline{\chi}}^{\mu\nu}} &=& 0\;, \label{6}\\
\int d^{4}x \frac{\delta\Sigma^{count}}{\delta\stw{\chi}^{\mu\nu}} &=& 0\;, \label{7}\\
\int d^{4}x \frac{\delta\Sigma^{count}}{\delta\stw{\overline{\psi}}^{\mu\nu}} &=& 0\;, \label{8}\\
\int d^{4}x \frac{\delta\Sigma^{count}}{\delta\stw{\psi}^{\mu\nu}} &=& 0\;, \label{9}\\
\int d^{4}x \frac{\delta\Sigma^{count}}{\delta\overline{\xi}^{\mu\nu}} &=& 0\;, \label{10}\\
{U^{(1)}}_{\sigma\lambda\mu\nu}(\Sigma^{count} ) &=& 0\;, \label{11}\\
{U^{(0)}}_{\sigma\lambda\mu\nu}(\Sigma^{count} ) &=& 0\;, \label{12}\\
{\widetilde{U}^{(0)}}_{\sigma\lambda\mu\nu}(\Sigma^{count} ) &=& 0\;, \label{13}\\
{U^{(2)}}_{\sigma\lambda\mu\nu}(\Sigma^{count} ) &=& 0\;, \label{14}\\
{\widetilde{U}^{(2)}}_{\sigma\lambda\mu\nu}(\Sigma^{count} ) &=& 0\; , \label{15}
\end{eqnarray}

where in (\ref{1}), ${\cal B}_{\Sigma}$ stands for the nilpotent linearized Slavnov-Taylor operator,
\begin{eqnarray}
{\cal B}_{\Sigma}&=&\int d^{4}x(
\frac{\delta\Sigma}{\delta\stw{A}_{\mu}}\frac{\delta}{\delta\stw{\Omega}^{\mu}}
+ \frac{\delta\Sigma}{\delta\stw{\Omega}^{\mu}}\frac{\delta}{\delta\stw{A}_{\mu}}
+ \frac{\delta\Sigma}{\delta\stw{c}}\frac{\delta}{\delta\stw{L}}
+ \frac{\delta\Sigma}{\delta\stw{L}}\frac{\delta}{\delta\stw{c}}
+ i\stw{b}\frac{\delta}{\delta\stw{\overline{c}}}\nonumber \\
&+&\frac{\delta\Sigma}{\delta\stw{\overline{u}}^{\mu\nu}}
\frac{\delta}{\delta\stw{\xi}_{\mu\nu}}
+ \frac{\delta\Sigma}{\delta\stw{\xi}_{\mu\nu}}
\frac{\delta}{\delta\stw{\overline{u}}^{\mu\nu}}
+\frac{\delta\Sigma}{\delta\stw{u}^{\mu\nu}}
\frac{\delta}{\delta\stw{\overline{\xi}}_{\mu\nu}}
+\frac{\delta\Sigma}{\delta\stw{\overline{\xi}}_{\mu\nu}}
\frac{\delta}{\delta\stw{u}^{\mu\nu}} \nonumber \\
&+& \frac{\delta\Sigma}{\delta\stw{\overline{v}}^{\mu\nu}}
\frac{\delta}{\delta\stw{B}_{\mu\nu}}
+ \frac{\delta\Sigma}{\delta\stw{B}_{\mu\nu}}
\frac{\delta}{\delta\stw{\overline{v}}^{\mu\nu}}
+ \frac{\delta\Sigma}{\delta\stw{v}^{\mu\nu}}
\frac{\delta}{\delta\stw{\overline{B}}_{\mu\nu}}
+ \frac{\delta\Sigma}{\delta\stw{\overline{B}}_{\mu\nu}}
\frac{\delta}{\delta\stw{v}^{\mu\nu}}\nonumber \\
&+& \frac{\delta\Sigma}{\delta\stw{\overline{P}}^{\mu\nu}}
\frac{\delta}{\delta\stw{\psi}_{\mu\nu}}
+ \frac{\delta\Sigma}{\delta\stw{\psi}_{\mu\nu}}
\frac{\delta}{\delta\stw{\overline{P}}^{\mu\nu}}
+ \frac{\delta\Sigma}{\delta\stw{P}^{\mu\nu}}
\frac{\delta}{\delta\stw{\overline{\psi}}_{\mu\nu}}
+ \frac{\delta\Sigma}{\delta\stw{\overline{\psi}}_{\mu\nu}}
\frac{\delta}{\delta\stw{P}^{\mu\nu}}
\nonumber \\
&+& \frac{\delta\Sigma}{\delta\stw{\overline{R}}^{\mu\nu}}
\frac{\delta}{\delta\stw{\chi}_{\mu\nu}}
+ \frac{\delta\Sigma}{\delta\stw{\chi}_{\mu\nu}}
\frac{\delta}{\delta\stw{\overline{R}}^{\mu\nu}}
+ \frac{\delta\Sigma}{\delta\stw{R}^{\mu\nu}}
\frac{\delta}{\delta\stw{\overline{\chi}}_{\mu\nu}}
+ \frac{\delta\Sigma}{\delta\stw{\overline{\chi}}_{\mu\nu}}
\frac{\delta}{\delta\stw{R}^{\mu\nu}}\nonumber \\
&+&
\frac{\delta\Sigma}{\delta J_{\sigma\lambda\alpha\beta}}
\frac{\delta}{\delta \overline{N}^{\sigma\lambda\alpha\beta}} +
\frac{\delta\Sigma}{\delta N^{\sigma\lambda\alpha\beta}}
\frac{\delta}{\delta \overline{J}_{\sigma\lambda\alpha\beta}}
+\frac{\delta\Sigma}{\delta\overline{J}_{\sigma\lambda\alpha\beta}}
\frac{\delta}{\delta N^{\sigma\lambda\alpha\beta}}
+\frac{\delta\Sigma}{\delta\overline{N}^{\sigma\lambda\alpha\beta}}
\frac{\delta}{\delta J_{\sigma\lambda\alpha\beta}}
 \nonumber \\
&+& \frac{\delta\Sigma}{\delta Q_{\sigma\lambda\alpha\beta}}
\frac{\delta}{\delta \overline{M}^{\sigma\lambda\alpha\beta}} +
\frac{\delta\Sigma}{\delta M^{\sigma\lambda\alpha\beta}}
\frac{\delta}{\delta \overline{Q}_{\sigma\lambda\alpha\beta}}
+ \frac{\delta\Sigma}{\delta\overline{Q}_{\sigma\lambda\alpha\beta}}
\frac{\delta}{\delta M^{\sigma\lambda\alpha\beta}}
+ \frac{\delta\Sigma}{\delta\overline{M}^{\sigma\lambda\alpha\beta}}
\frac{\delta}{\delta Q_{\sigma\lambda\alpha\beta}}
)\;.\nonumber \\
{\cal B}^{2}_{\Sigma} &=& 0 ,
\label{slav}
\end{eqnarray}
and in (\ref{4}),
\begin{eqnarray}
D_{\Sigma} &=& \int d^{4}x \lbrace c\frac{\delta }{\delta\stw{\overline{c}}} -
i \frac{\delta\Sigma}{\delta b}\frac{\delta }{\delta L} - i \frac{\delta\Sigma}{\delta L}\frac{\delta }{\delta b}\rbrace = 0.
\label{Dsl2r}
\end{eqnarray}

The first constraint (\ref{1}) together with (\ref{slav}), establishes a cohomological problem for
the operator ${\cal B}_\Sigma$ and its solution is given by \cite{livro}
\begin{equation}
\Sigma^{count}=\frac{a_0}{4}\int d^{4}x\stw{F}_{\mu\nu}\ast\stw{F}^{\mu\nu}
+\Delta^{\left(  0\right)  }, \hspace{1cm}
\Delta^{\left(  0\right)  } = {\cal B}_{\Sigma}\Delta^{\left(  -1\right)  }\;,\label{count00}
\end{equation}
where $\Delta^{(0)}$ is a local integrated polynomial in all
fields and sources, with ultra-violet dimension up-bounded by
four, ghost number zero and vanishing Q charge. The other Ward
identities (\ref{2}) to (\ref{15}) will give constraints to
$\Delta^{(0)}$. In the first place, equations (\ref{2}) and
(\ref{3}) state that $b$ cannot be used in its construction, and
that the source $\Omega_{\mu}$ and the antighost $\overline{c}$
can only appear in the combination $\Omega_{\mu} +
\partial_{\mu}\overline{c}$. Equations (\ref{4}) and (\ref{5}) are
also typical of gauge theories and fix coefficients of
counterterms already present at the original action. Now, it is of
fundamental importance to note that, due to equations
(\ref{6},\ref{7},\ref{8},\ref{9},\ref{10}), the fields
$\overline{\chi}$,$\chi$,$\overline{\psi}$,$\psi$ and
$\overline{\xi}$ only appear directly derivated or inside Moyal
comutators (anticomutators). In fact, this is also valid for all
BRST sources in the theory, which obey similar equations.

Now, if we concentrate ourselves on contributions that can damage
the IR equilibrium established in (\ref{snl}), we must look for
counterterms that may modify the gauge propagation coming from
this action. The first one that comes to mind is
\begin{equation}
\int d^{4}x ( \overline{B}_{\mu\nu}B^{\mu\nu}  - \overline{\xi}_{\mu\nu}\xi^{\mu\nu}), \label{BBbar}
\end{equation}
which, although being allowed by all the remaining Ward
identities, is avoided by the eq. (\ref{10}).

Another possible counterterm which deserves special attention is
\begin{equation}
\int d^{4}x ( \overline{B}_{\mu\nu}D^{2}B^{\mu\nu}  - \overline{\xi}_{\mu\nu}D^{2}\xi^{\mu\nu}), \label{BDBbar}
\end{equation}
 which is not allowed explicitly by the identity (\ref{14}).

There is also the element
\begin{equation}
a\int d^{4}x (\overline{\chi}_{\mu\nu}D^{2}\chi^{\mu\nu}  -
\overline{\psi}_{\mu\nu}D^{2}\psi^{\mu\nu}), \label{chiDchibar}
\end{equation}
which is in fact allowed by all the symmetries. This counterterm,
not originally present in the localized action (\ref{acao}),
changes the propagator to
\begin{equation}
\left\langle A(k)_\mu A_\nu(-k)\right\rangle=\left(\delta_{\mu\nu}
-\frac{k_\mu k_\nu}{k^2}\right)\frac{k^2 }{k^{4} + a\gamma^{2}k^2
+ \gamma^{4} }\; .\label{d00}
\end{equation}

This form of the propagator still means that the IR ambiguity is
eliminated, as one can see by rewriting the new non-local theory
in the presence of the term (\ref{chiDchibar}).

Now, we should not forget that the original theory, eq.
(\ref{localghost}), that we are studying is actually a limit of
the larger theory described by (\ref{acao}), when the sources
reach their physical values (\ref{pval}). Then, there is still a
class of possible counterterms that eventually can change the
propagator (or the non-local action) but that appear in the
larger theory as 4-point divergent contributions. In particular,
we have that the element

\begin{equation}
\alpha\int d^{4}x (\lbrace
\overline{J}_{\alpha\beta\sigma\lambda}\stackrel{*}{,}J^{\alpha\beta\sigma\lambda}\rbrace
\lbrace \overline{B}_{\mu\nu}\stackrel{*}{,}B^{\mu\nu}\rbrace
-\lbrace
\overline{Q}_{\alpha\beta\sigma\lambda}\stackrel{*}{,}J^{\alpha\beta\sigma\lambda}\rbrace
\lbrace \overline{B}_{\mu\nu}\stackrel{*}{,}\xi^{\mu\nu}\rbrace )
\label{dbb}
\end{equation}
is also allowed by all the symmetries from (\ref{1}) to
(\ref{15}).

 These two terms are then responsible for a gauge
propagator modified in relation to that in (\ref{ncmaxprop}). When
the sources $J,\overline{J},Q$ and $\overline{Q}$ are set to their
physical values, the propagator for the gauge field takes the
general form:
\begin{eqnarray}
\left\langle A(k)_\mu A_\nu(-k)\right\rangle &=&
\left(\delta_{\mu\nu}
-\frac{k_\mu k_\nu}{k^2}\right)\frac{\Xi^{2}}{k^{2} (\Xi^{2} +\gamma^{2}\Pi )}\nonumber \\
\Xi &=& k^{4} - \alpha a\gamma^{2}k^{2} + \alpha\gamma^{4} \nonumber \\
\Pi &=& a k^{6} + (1- \alpha a^{2})\gamma^{2}k^{4} -\alpha
\gamma^{6}\; .\label{d11}
\end{eqnarray}
This means that the inclusion of all counterterms
in the starting classical action will in the
end destroy the IR mechanism proposed in (\ref{snl}).
Unfortunately , the element (\ref{dbb}) seems to be actually found
in explicit graphic calculations and it is clear that only the
case $\alpha=0$ would correspond to a well behaved propagator.

It is important to mention that the cohomological analysis
extended to the noncommutative space is constrained by the Ward
identities of the action. If another Ward identity is observed,
this constraint may reduce the number of counterterms. One example
is the counterterm responsible for the mass $a$ that is apparently
not required at one loop calculations.

It should be stressed that although the choice of $a$ and $\alpha
$ different from $0$ at tree level would give rise to a very
different type of propagator, the ultraviolet behavior is exactly
$\frac{1}{k^{2}}$. With an adequate choice for these parameters,
it is possible that the propagator satisfy the Wilson criterium
for confinement \cite{Fisch,coulomb}. The Wilson criterium and the
loss of positivity are interpreted as a sign of confinement
\cite{zwanziger,alkofer,mendes,dudal}. This would possibly mean
that confining phases can be expected in noncommutative gauge
theories. In such case the physical excitations are not associated
to the fundamental fields and only condensates of fields are good
candidates to physical states of the model \cite{dudal}. Another
important point is that in this context the Wick rotation is not
allowed in general. But there is still the possibility that the
correlator between two condensates have a massive particle pole.
These correlators admit Wick rotation and can be associated to
observable physical states in Minkowsky space \cite{dudal}.
These observations may be useful for a future understanding
of the nature of noncommutative gauge theories.

\section{Conclusion}

We saw along this work how a nonlocal mechanism as that in
equation (\ref{snl}), that classically can cure the infrared
problem of the 2-point function of a noncommutative Maxwell theory
is not ultraviolet stable.

In the development of this algebraic proof, we followed the
approach used by \cite{zwanziger} , and more recently improved by
Sorella and Baulieu \cite{baulieu} , to the study of the BRST
quantization of the nonlocal action coming from  Gribov's
observations on the infrared properties of gauge theories. We
understand that, if in the usual commutative space the use of
nonlocal actions is an alternative option to the study of the
infrared regime, on the other hand, in the noncommutative case
this seems to be the inevitable path to solve the intrinsic
problem of the IR/UV mix.

As a final comment, we would like to point out a recent proposal simplifying
(\ref{snl}) in order to avoid the quantum generation of counterterms as (\ref{dbb}),
but still preserving the IR match for the 2-point function \cite{U*(1)}

\section*{Acknowledgments}

We thank Paolo Castorina for valuable discussions about 
symmetry breaking in noncommutative theories.
The work of V.E.R.Lemes was supported by Conselho Nacional de
Desenvolvimento Cient\'{\i}fico e Tecnol\'{o}gico (CNPq-Brazil),
Funda{\c{c}}{\~{a}}o de Amparo \`a Pesquisa do Estado do Rio de
Janeiro (Faperj) and SR2-UERJ . O.S.Ventura was supported by
Conselho Nacional de Desenvolvimento Cient\'{\i}fico e
Tecnol\'{o}gico (CNPq-Brazil) and also at the beginning of this
work by FUNCEFETES and Pro-Reitoria de Pesquisa e Pos-graduacao at
Ifes.

\end{document}